\documentclass[12pt]{article}
\usepackage{amssymb}
\usepackage{amsmath}
\usepackage{amsfonts}
\usepackage{amsthm}

\font\frak=eufm10 scaled\magstep1  
\font\black=msbm10 scaled\magstep1 \font\bigblack=msbm10 scaled\magstep 2 \font\bbigblack=msbm10
scaled\magstep3

\def\goth #1{\hbox{{\frak #1}}}

\def\bigfield #1{\hbox{{\bigblack #1}}}
\def\bbigfield #1{\hbox{{\bbigblack #1}}}
\def\Bbb #1{\hbox{{\black #1}}}
\def\v #1{\vert #1\vert}             %Para denotar elgrado de #1

\def\m #1 #2{(-1)^{{\v #1} {\v #2}}} %Para denotar el signo (-1)^...
\def\pd#1#2{\frac{\partial#1}{\partial#2}}

\def\<#1>{\langle#1\rangle}        %  una forma bilineal <x,a>
\def\>#1{{\bf #1}}                %  notacin para vectores
\def\f(#1,#2){\frac{#1}{#2}}

\def\dt2#1{\frac{d^2 #1}{dt^2}}

\def\ea{\varepsilon_a}

\def\R{{\hbox{{\field R}}}}             %%real numbers (Pepin)
\def\big R{{\hbox{{\bigfield R}}}}
\def\bbig R{{\hbox{{\bbigfield R}}}}
\def\dim{\hbox{{\rm dim}}}        %%several definitions

\def\R{\mathbb{R}}
\def\ba{\begin{eqnarray}}
\def\ea{\end{eqnarray}}
\def\be{\begin{equation}}
\def\ee{\end{equation}}

\def\R{\Bbb R}

\def\goth #1{\hbox{{\frak #1}}}
\def\<#1>{\langle#1\rangle}

\def\pd#1#2{\frac{\partial#1}{\partial#2}}

\def\Diff{\operatorname{Diff}}
\def\di{\bigstar}
\newcommand{\bea}{\begin{eqnarray}}
\newcommand{\eea}{\end{eqnarray}}

\def\frac#1#2{{#1\over#2}} \def\pd#1#2{\frac{\partial#1}{\partial#2}}
\newtheorem{prop}{Proposici\'on}[section]
\newtheorem{note}[prop]{Note}

\theoremstyle{plain}
\newtheorem{theorem}{Theorem}

\newtheorem{proposition}{Proposition}
\newtheorem{definition}{Definition}

\newtheorem{remark}{Remark}

\def\R{\Bbb R}

\font\frak=eufm10 scaled\magstep1

\def\goth #1{\hbox{{\frak #1}}}
\def\<#1>{\langle#1\rangle}

\begin{document}

\centerline{\Large {\bf Quasi--Lie schemes: theory  and applications}} \vskip 0.75cm

\centerline{ Jos\'e F. Cari\~nena$^{\dagger}$, Janusz Grabowski$^{\ddagger}$ and Javier de
Lucas$^{\dagger}{}^{\ddagger}$} \vskip 0.5cm

\centerline{$^{\dagger}$Departamento de  F\'{\i}sica Te\'orica, Universidad de Zaragoza,}
\smallskip
\centerline{50009 Zaragoza, Spain.}
\medskip
\centerline{$^{\ddagger}$Institute of Mathematics, Polish Academy of Sciences,}
\smallskip
\centerline{ul. \'Sniadeckich 8, P.O.~Box 21, 00-956 Warszawa, Poland}
\smallskip
\centerline{}
\medskip

\vskip 1cm

\begin{abstract}
A powerful method to solve nonlinear first-order ordinary differential equations, which is based on a
geometrical understanding of the corresponding dynamics of the so-called Lie systems, is developed. This
method enables us not only to solve some of these equations, but also gives geometrical explanations for some,
already known, {\it ad hoc} methods of dealing with such problems.



\end{abstract}

\section{Introduction}
\qquad Systems of nonautonomous first-order ordinary differential equations appear often in Mathematics,
Physics, Chemistry and Engineering and therefore methods to
 solve  them  and analyse their
properties are specially interesting because they allow us to understand many important problems in these
various fields.

A special kind of these systems, the so-called {\it Lie systems} (or {\it Lie-Scheffers systems})
\cite{LS}-\cite{PW}, has recently been analysed in many papers \cite{CRL07a}-\cite{CL08}. An important
property of such systems  is that they admit a certain set of time-dependent changes of variables which
transforms each Lie system into a new one \cite{CarRam05b}. Such changes can be used to transform a given Lie
system into an easily integrable one, i.e. into a Lie system related to a solvable Lie algebra of vector
fields. In all these cases, we can obtain constants of the motion, integrability conditions or even solutions
\cite{CL08}.

This transformation property is not only valid  for Lie systems but it still holds for a more general set of
systems of differential equations. In order to generalise this property we develop the  concept of a {\it
quasi-Lie system}, i.e. a system of nonautonomous differential equations accompanied with a flow of
diffeomorphisms transforming it into a Lie system. We must stress that quasi-Lie systems are not equivalent to
Lie systems in the trivial sense: the transformations we use are time-dependent and we might get
time-dependent superposition rules for them, meanwhile, Lie systems are related to superposition rules which
are time-independent. Furthermore we define {\it quasi-Lie schemes}. A scheme is a procedure that, in some
cases, enables us to transform a given nonautonomous system of differential equations into a Lie system. This
scheme, when applied to Lie systems, gets back to the well-known transformations of Lie systems, see
\cite{CLR07b}. However, many other important systems of first-order differential equations can be studied by
means of this new method. In this sense the method of quasi-Lie schemes may be viewed as an abstract
unification of many {\it ad hoc} developed integration methods existing in the literature.

The aim of this paper is to introduce and to analyse the properties of quasi-Lie schemes and to illustrate the
theory by developing some interesting examples. More specifically, in order to the paper be self-contained,
Section 2 is devoted to review the theory of time-dependent vector fields. Then we summarise in Section 3 some
properties of Lie systems and in Section 4 we study the quasi-Lie scheme concept which may be used to relate
certain sort of systems of differential equations to Lie systems.   Such quasi-Lie schemes are applied in
Section 5 to analyse some interesting systems of differential equations. Some well-known differential equations are analysed: dissipative Milne--Pinney equations, nonlinear oscillator and Emden equations  in order to recover from an unified point of view some of their properties.  The fields of applications of these differential equations is very broad, i.e. just the Emden equation appears in Mathematical Physics \cite{ZP94}, Theoretical Physics \cite{MP99}, Astronomy \cite{MH06},
 Astrophysics \cite{Li96} and the review by Wong \cite{Wong76a} contains about 140 references. We also study a time-dependent dissipative Mathews--Lakshmanan oscillator and we provide a, as far as we know, new time-dependent constant of the motion. Finally, in Section 6, we sum up the
conclusions of our paper and give an outlook of possible problems to be studied by means of the methods here
developed.

\section{Generalised flows and time-dependent vector fields}

\qquad A nonautonomous system of first-order ordinary differential equations in a manifold $N$ is represented
by a time-dependent vector field $X=X(t,x)$ on such a manifold. On a non-compact manifold, the vector field
$X_t(x)=X(t,x)$, for a fixed $t$, is generally not defined globally, but it is well defined on a neighbourhood
of each point $x_0\in N$ for sufficiently small $t$. It is convenient to add the time $t$ to the manifold and
to consider the {\it autonomisation} of our system, i.e. the vector field
$$\bar{X}(t,x)=\pd{}{t}+X(t,x)\,,
$$
defined on a neighbourhood $N^X$ of $\{0\}\times N$ in $\mathbb{R}\times N$. The vector field $X_t$ is then
defined on the open set of $N$,
$$ N_t^X=\{x_0\in N\mid (t,x_0)\in N^X\}\,,$$
 for all $t\in \mathbb{R}$. If $N_t^X=N$ for all $t\in
\mathbb{R}$, we speak about a {\it global time-dependent vector field}. The  system of differential equations
associated with the time-dependent vector field $X(t,x)$ is written in local coordinates
$$\frac{dx^i}{dt}=X^i(t,x)\,,\qquad i=1,\ldots,n=\dim\, N,
$$
where $X(t,x)=\sum_{i=1}^nX^i(t,x)\partial/\partial{x^i}$ is defined locally on the manifold for sufficiently
small $t$.

A solution of this system is represented by a curve $s\mapsto \gamma(s)$ in $N$ (integral curve) whose tangent
vector $\dot \gamma$ at $t$, so at the point $\gamma(t)$ of the manifold, equals $ X(t, \gamma(t))$. In other
words,
\begin{equation}\label{e1}
\dot \gamma(t)= X(t, \gamma(t)).
\end{equation}
It is well-known that, at least for smooth $X$ which we work with, for each $x_0$ there is a unique maximal
solution $\gamma_X^{x_0}(t)$  of system (\ref{e1}) with the initial value $x_0$, i.e. satisfying
$\gamma_X^{x_0}(0)=x_0$. This solution is defined at least for $t$'s from a neighbourhood of $0$. In case
$\gamma_X^{x_0}(t)$ is defined for all $t\in \mathbb{R}$, we speak about a {\it global time solution}.

The collection of all maximal solutions of the system (\ref{e1}) gives rise to a (local) generalised flow
$g^X$ on $N$. By a {\it generalised flow} $g$ on $N$ we understand a smooth time-dependent family $g_t$ of
 local diffeomorphisms on $N$, $g_t(x)=g(t,x)$, such that
$g_0=\text{id}_N$. More precisely, $g$ is a smooth map from a neighbourhood $N^g$ of $\{0\}\times N$ in
$\mathbb{R}\times N$ into $N$, such that $g_t$ maps diffeomorphically the open submanifold $N^g_t=\{x_0\in
N\mid (t,x_0)\in N^g\}$ onto its image, and $g_0=\text{id}_N$. Again, for each $x_0\in N$ there is a
neighbourhood $U_{x_0}$ of $x_0$ in $N$ and $\epsilon>0$ such that $g_t$ is defined on $U_{x_0}$ for
$t\in(-\epsilon,\epsilon)$ and maps $U_{x_0}$ diffeomorphically onto $g_t(U_{x_0})$.

If $N_t^g=N$ for all $t\in \mathbb{R}$, we speak about a {\it global generalised flow}. In this case $g:t\in
\R\mapsto g_t\in\Diff(N)$ may be viewed as a smooth curve in the diffeomorphism group $\Diff(N)$ with
$g_0=\text{id}_N$.

Here is also convenient to {\it autonomise} the generalised flow $g$ extending it to a single local
diffeomorphism
\begin{equation}\label{e2} \bar{g}(t,x)=(t,g(t,x))
\end{equation}
defined on the neighbourhood $N^g$ of $\{0\}\times N$ in $\mathbb{R}\times N$. The generalised flow $g^X$
induced by the time-dependent vector field $X$ is defined by
\begin{equation}\label{e3}
g^X(t,x_0)=\gamma_X^{x_0}(t)\,.
\end{equation}
Note that, for $g=g^X$, equation (\ref{e3}) can be rewritten in the form:
\begin{equation}\label{e4}
X_t= X(t,x)=\dot g_t\circ g_t^{-1}\,.
\end{equation}
In the above formula, we understood $ X_t$ and $\dot g_t$ as maps from $N$ into $TN$, where $\dot g_t(x)$ is
the vector tangent to the curve $s\mapsto g(s,x)$ at $g(t,x)$. Of course, the composition $\dot g_t\circ
g_t^{-1}$, called sometimes the {\it right-logarithmic derivative} of $t\mapsto g_t$, is defined only for
those points $x_0\in N$ for which it makes sense. But it is always the case locally for sufficiently small
$t$.

Let us observe that equation (\ref{e4}) defines, in fact, a one-to-one correspondence between generalised
flows and time-dependent vector fields modulo the observation that the domains of $\dot g_t\circ g_t^{-1}$ and
$ X_t$ need not coincide. In any case, however, $\dot g_t\circ g_t^{-1}$ and $ X_t$ coincide in a
neighbourhood of any point for sufficiently small $t$. One can simply say that the {\it germs} of $X$ and
$\dot g_t\circ g_t^{-1}$ coincide, where the germ in our context is understood as the class of corresponding
objects that coincide on a neighbourhood of $\{0\}\times N$ in $\mathbb{R}\times N$.

Indeed, for a given $g$, the corresponding time-dependent vector field is defined by (\ref{e4}). Conversely,
for a given $X$, the equation (\ref{e4}) determines the germ of the generalised flow $g(t,x)$ uniquely, as for
each $x=x_0$ and for small $t$ equation (\ref{e4}) implies that $t\mapsto g(t,x_0)$ is the solution of the
system defined by $X$ with the initial value $x_0$. In this way we get the following
\begin{theorem}\label{t1} Equation (\ref{e4}) defines a
one-to-one correspondence between the germs of generalised flows and the germs of time-dependent vector fields
on $N$. For compact $N$, this correspondence reduces to a one-to-one correspondence between global
time-dependent vector fields and global generalised flows.
\end{theorem}
\noindent Any two generalised flows $g$ and $h$ can be composed: by definition $(g\circ h)_t=g_t\circ h_t$,
where, as usual, we view $g_t\circ h_t$ as a local diffeomorphism defined for points for which the composition
makes sense. It is important that in a neighbourhood of any point it really makes sense for sufficiently small
$t$. As generalised flows correspond to time-dependent vector fields, this gives rise to an action  of a
generalised flow $h$ on a time-dependent vector field $X$, giving rise to $h_\di X$ defined  by the equation
\begin{equation}
g^{h_\di X}=h\circ g^X\,. \label{e5}
\end{equation}
To obtain a more explicit form of this action, let us observe that
$$
(h_\di X)_t=\frac{d(h\circ g^X)_t}{dt}\circ (h\circ g^X)_t^{-1}=\left( \dot h_t\circ g_t^X+Dh_t(\dot g_t^X)
\right)\circ (g^X)_t^{-1}\circ h_t^{-1},
$$
and therefore
$$(h_\di X)_t=\dot
h_t\circ h_t^{-1}+Dh_t\left (\dot g_t^X\circ (g^X)_t^{-1}\right)\circ h_t^{-1},
$$
i.e.
\begin{equation}\label{e6}
(h_\di X)_t=\dot h_t\circ h_t^{-1}+(h_t)_*(X_t)\,,
\end{equation}
where $(h_t)_*$ is the standard action of diffeomorphisms on vector fields. In a slightly different form, this
can be written as an action of time-dependent vector fields on time-dependent vector fields:
\begin{equation}\label{e7} (g^Y_\di X)_t=Y_t+(g_t^Y)_*(X_t)\,.
\end{equation}
For global time-dependent vector fields on compact manifolds the latter defines a group structure in global
time-dependent vector fields. This is an infinite-dimensional analog of a group structure on paths in a
finite-dimensional Lie algebra which has been used as a source for a nice construction of the corresponding
Lie group in \cite{DK}. \noindent Since every generalised flow has the inverse, $(g^{-1})_t=(g_t)^{-1}$, so
generalised flows, or better to say, the corresponding germs, form a group and the formula (\ref{e7}) allows
us to compute the time-dependent vector field (right-logarithmic derivative) $X_t^{-1}$ associated with the
inverse. It is the time-dependent vector field
\begin{equation}
X_t^{-1}=-(g^X_t)^{-1}_*(X_t)\,.
\end{equation}
For time-independent vector fields $X_t=X_0$ for all $t$  we have $(g^X_t)_* X=X$ and also we get the
well-known formula
$$X^{-1}=-X\,.$$
Note that, by definition, the integral curves of $h_\di X$ are of the form $h_t(\gamma(t))$, where $\gamma(t)$
are integral curves of $X$. We can summarise our observation as follows.
\begin{theorem}\label{t2} The equation (\ref{e6}) defines a natural action
of  generalised flows  on  time-dependent vector fields. This action is a group action in the sense that
$$(g\circ h)_\di X=g_\di(h_\di X)\,.
$$
The integral curves of $h_\di X$ are of the form $h_t(\gamma(t))$, for $\gamma(t)$ being an arbitrary integral
curve for $X$.
\end{theorem}
\noindent The above action of generalised flows on time-dependent vector fields can be also defined in an
elegant way by means of the corresponding autonomisations. It is namely easy to check the following.
\begin{theorem}\label{t3} For any generalised flow $h$ and any
time-dependent vector field $X$ on a manifold $N$, the standard action $\bar{h}_*\bar{X}$ of the
diffeomorphism $\bar{h}$, being the autonomisation of $h$, on the vector field $\bar{X}$, being the
autonomisation of $X$, is the autonomisation of the time-dependent vector field $h_\di X$:
$$\bar{h}_*\bar{X}=\overline{h_\di X}\,.$$
\end{theorem}

\section{Lie systems and superposition rules}
\qquad The conditions for the system determined by a time-dependent vector field $X(t,x)$ on a manifold $N$
ensuring that it admits a {\it superposition rule}, i.e. that there exists, at least locally for an open $U\subset N^{m}\times\R^n$, a map $\Phi:U\to N$, $x=\Phi(x_{(1)}, \ldots,x_{(m)};k_1,\ldots,k_n)$, such that its
general solution can be written as $$x(t)=\Phi(x_{(1)}(t), \ldots,x_{(m)}(t);k_1,\ldots,k_n),$$ where
$\{x_{(a)}(t)\mid a=1,\ldots,m\}$ is any family of particular solutions `in general position' and
$k=(k_1,\ldots,k_n)$ is a set of $n$ arbitrary  constants, were studied by S. Lie \cite{LS}. Let us stress
that the superposition function $\Phi$ is time-independent.

The necessary and sufficient conditions say that the associated time-dependent vector field $X$ can be written
as a linear combination
\begin{equation}
X_t=\sum_{\alpha =1}^r b_\alpha(t)\, X_{(\alpha )}\ , \label{Lievf}
\end{equation}
such that the vector fields
 $\{X_{(\alpha)}\mid \alpha=1,\ldots,r\}$ generate a
 finite-dimensional  real
 Lie algebra, the so-called Vessiot-Guldberg Lie algebra. The latter means that there exist
 $r^3$ real numbers $c_{\alpha\beta\gamma}$, such that
\begin{equation*}
[X_{(\alpha)},X_{(\beta)}]=\sum_{\gamma=1}^r c_{\alpha\beta\gamma }X_{(\gamma)}\ ,\qquad
\alpha,\beta=1,\ldots,r\,.
\end{equation*}
Linear systems are particular instances of Lie systems associated with a Vessiot-Guldberg Lie algebra,
isomorphic to the Lie algebra $\goth{gl}(n,\mathbb{R})$, and $m=n$ in the homogeneous case, or the
corresponding affine algebra and $m=n+1$ in the inhomogeneous one. The Riccati equation is another example for
which $X_t=b_0(t)Y_0+b_1(t)Y_1+b_2(t)Y_2$ with
\begin{equation*}
Y_0=\pd{}x\,,\qquad Y_1=x\,\pd{}x\,,\qquad Y_2=x^2\,\pd{}x\,,
\end{equation*}
% and commutation relations
% $$
% [Y_0,Y_1]=Y_0,\quad [Y_1,Y_2]=Y_2,\quad [Y_0,Y_2]=2Y_1,
% $$
closing on a Vessiot-Guldberg Lie algebra isomorphic to the Lie algebra $\goth{sl}(2,\mathbb{R})$, see
\cite{CLR07b,CarRam}.
There is an action $\Phi_{\rm Ric}:SL(2,\mathbb{R})\times \bar{\mathbb{R}}\rightarrow
\bar{\mathbb{R}}$ of the Lie group $SL(2,R)$ on $\bar{\mathbb{R}}\equiv \mathbb{R}\cup \infty$ \cite{CLR07b} such that the fundamental vector fields of this action are linear combinations with real coefficients of the
vector fields $Y_0, Y_1$ and $Y_2$.

Another relevant example of a Lie system is given by a  time-dependent right-invariant vector field in a Lie
group $G$. If   $\{{\rm a}_1,\ldots,{\rm a}_r\}$ is a basis of $T_eG$ and $X^{\tt R}_\alpha$  are the
corresponding
 right-invariant vector fields, $X^{\tt R}_\alpha(g)=(R_g)_*{\rm a}_\alpha$, then the
 time-dependent right-invariant vector field
$$X_t=-\sum_{\alpha=1}^r b_\alpha(t)X^{\tt R}_\alpha,$$
defines a Lie system in $G$ whose integral curves are  solutions of the system $\dot g=-\sum_{\alpha=1}^r
b_\alpha(t)\,X^{\tt R}_\alpha(g)$. Applying $(R_{g^{-1}(t)})_*$ to both sides, we see that $g(t)$ satisfies
\begin{equation}\label{eqLG}
(R_{g^{-1}(t)})_*\dot g(t)\,=-\sum_{\alpha=1}^r b_\alpha(t){\rm a}_\alpha\in T_eG \ .
\end{equation}
Right-invariance means that it is enough to know one solution, for instance the one starting from the neutral
element $e$, to know all the solutions of the equation with any initial condition, i.e. we obtain the solution
$g'(t)$  with the initial condition $g'(0)=g_0$ as $R_{g(0)}g(t)$. A generalisation of the method used by Wei
and Norman for linear systems \cite{WN} is very useful in solving such equations and furthermore there exist
reduction techniques that can also be used \cite{CarRamGra}. Finally, as right-invariant vector fields $X^{\tt
R}$ project onto the fundamental vector fields in each homogeneous space of $G$, the solution of (\ref{eqLG})
allows us to find the general solution for the corresponding Lie system in each homogeneous space. Conversely,
the knowledge of particular solutions of the associated system in a homogeneous space gives us a method for
reducing the problem to the corresponding isotropy group \cite{CarRamGra}. This equation is also important
because any Lie system described by a time-dependent vector field on a manifold
 $N$, like (\ref{Lievf}), where the vector fields are complete
  and  satisfy the same
commutation
 relations as the basis $\{{\rm a}_1,\ldots,{\rm a}_r\}$, determines an action
$\Phi_{\rm LieG}:G\times N\rightarrow N$ such that the  vector field $X_\alpha$ is the fundamental vector
field corresponding to  ${\rm a}_\alpha$, and moreover, the integral curves for the time-dependent vector
field are obtained from the solutions of  equation (\ref{eqLG}). More explicitly,
 the general solution of the given  Lie system is
 $x(t)=\Phi_{\rm LieG}(g(t),x_0)$, where $x_0$ is
 the initial condition of the solution and $g(t)$ is the solution of equation (\ref{eqLG}) with $g(0)=e$.

The search for the number $m$ of solutions and the superposition function $\Phi$ has recently been studied
from a geometric perspective \cite{CGM07}. Essentially, we should consider  `diagonal prolongations' to
$N^{m+1}$,  $\widetilde X(x_{(0)}, \ldots, x_{(m)},t)$, of the time-dependent  vector field
$X(t,x)=\sum_{i=1}^nX^i(t,x)\,\partial/\partial x^i$, given by
\[
\widetilde X(x_{(0)}, \ldots,x_{(m)},t)=\sum_{a=0}^mX_a(x_{(a)},t)\,,\qquad t\in {\mathbb{R}}\,,
\]
where $X_{a}(x_{(a)},t)=\sum_{i=1}^nX^i(x_{(a)},t)\,{\partial}/{\partial x^i_{(a)}}$, such that the extended
system admits $n$ independent constants of the motion, which define in an implicit way the superposition
function.

\section{Quasi-Lie systems and schemes}\label{Theory}

\begin{definition}{\rm By a {\it quasi-Lie system} we understand a pair $(X,g)$ consisting of a time-dependent vector field
$X$ on a manifold $N$ (the {\it system}) and a generalised flow $g$ on $N$ (the {\it control}) such that
$g_\di X$ is a Lie system.}
\end{definition}
\noindent Since for the Lie system $g_\di X$ we are able to produce the general solution out of a number of
known particular solutions, the knowledge of the control makes a similar procedure for our initial system
possible. Indeed, let $\Phi=\Phi(x_{1},\dots,x_{m},k_1,\dots,k_n)$ be a superposition function for the Lie system
$g_\di X$, so that, knowing $m$ solutions $\bar x_{(1)},\ldots,\bar x_{(m)}$ of $g_\di X$, we can derive the
general solution of the form
$$\bar x_{(0)}=\Phi(\bar x_{(1)},\ldots,\bar  x_{(m)},k_1,\ldots, k_n)\,.
$$
If we now know $m$ independent solutions, $x_{(1)},\ldots, x_{(m)}$ of $X$, then, according to Theorem
\ref{t3}, $\bar x_{a}(t)=g_t(x_{a}(t))$ are solutions of $g_\di X$, producing a general solution of $g_\di X$
in the form $\Phi(\bar x_{(1)},\ldots,\bar x_{(m)},k_1,\ldots, k_n)$. It is now clear that
\begin{equation}\label{e8}
x_{(0)}(t)=g_t^{-1}\circ \Phi(g_t(x_{(1)}(t)),\ldots,g_t(x_{(m)}(t)),k_1,\ldots,k_n)
\end{equation}
is a general solution of $X$. In this way we have obtained a {\it time-dependent superposition rule} for the
system $X$. We can summarise the above considerations as follows.
\begin{theorem}\label{t4} Any quasi-Lie system $(X,g)$ admits a
time-dependent superposition rule of the form (\ref{e8}), where $\Phi$ is a superposition function for the Lie
system $g_\di X$.
\end{theorem}
\noindent Of course, the above time-dependent superposition rule is practically meaningful in finding the
general solution of a system $X$ only if the generalised flow $g$ is explicitly known. An alternative abstract
definition of a quasi-Lie system as a time-dependent vector field $X$ for which there exists a generalised
flow $g$ such that $g_\di X$ is a Lie system does not have much sense, as every $X$ would be a quasi-Lie
system
 in this
context. For instance, given a time-dependent vector field $X$, the pair $(X, (g^X)^{-1})$ is a quasi-Lie
system because $(g^X)_t^{-1}\circ g_t^X={\rm id}_N$, thus $(g^X)^{-1}_\di X=0$, which is a Lie system
trivially. On the other hand, finding $(g^X)^{-1}$ is nothing but solving our system $X$ completely, so we
just reduce to our original problem. In practice, it is therefore crucial that the control $g$ comes from a
system which can be integrated effectively. There are, however, many cases when our procedure works well and
provides a geometrical interpretation of many, originally developed {\it ad hoc}, methods of integration.
Consider, for instance, the following scheme that can lead to `nice' quasi-Lie systems.

Take a finite-dimensional real vector space $V$ of vector fields on $N$ and consider the family of all
time-dependent vector fields $X$ on $N$ such that $X_t$ belongs to $V$ on its domain, i.e. $X_t\in
V_{|N^X_t}$. We will say that these are time-dependent vector fields taking values in $V$. The time-dependent
vector fields taking values in $V$ depend on a finite family of control functions. For example, take a basis
$\{X_1,\dots,X_r\}$ of $V$ and consider a general time-dependent system with values in $V$ determined by
$b=b(t)=(b_1(t),\dots,b_r(t))$ as
$$(X^{b})_t=b_j(t)X_j\,.$$
On the other hand, the nonautonomous systems of differential equations associated with $X\in V|_{N_t^X}$ are
not Lie systems in general, if $V$ is not a Lie algebra itself. If we have additionally a finitely
parametrised family of local diffeomorphism, say $\underline{g}=\underline{g}(a_1,\dots,a_k)$, then any curve
$a=a(t)=(a_1(t),\dots,a_k(t))$ in the control parameters, defined for small $t$, gives rise to a generalised
flow $g^{a}_t=\underline{g}(a(t))$. Let us assume additionally that there is a Lie algebra $V_0$ of vector
fields contained in $V$. We can look for control functions $a(t)$ such that for certain $b(t)$ we get that
$g^{a}_\di X^{b}$ has values in $V_0$ for each time $t$. Let us denote this as
\begin{equation}\label{e9}g^{a}_\di X^{b}\in V_0.
\end{equation}
This choice of control functions makes $(X^{b},g^{a})$ into a quasi-Lie system, so we get time-dependent
superposition rules for the corresponding systems $X^b$.

Let us observe that in the case when all the generalised flows $g^a$ preserve $V$, i.e. for each
time-dependent vector field $X^b\in V$ also $g^a_\di X^b\in V$, the inclusion (\ref{e9}) becomes a
differential equation for the control functions $a(t)$ in terms of the functions $b(t)$. This situation is not
so rare as it may seem to be at the first sight. Suppose, for instance, that we find a Lie algebra $W\subset
V$ such that $[W,V]\subset V$ and that the time-dependent vector fields with values in $W$ can be effectively
integrated to generalised flows. In this case, any time-dependent vector field $Y^{a}$ with values in $W$
gives rise to a generalised flow $g^{a}$ which, in view of transformation rule (\ref{e7}), preserves the set
of time-dependent vector fields with values in $V$. For each $b=b(t)$ the inclusion (\ref{e9}) becomes
therefore a differential equation for the control function $a=a(t)$ which often can be effectively solved.
\begin{definition}\label{QLscheme}{\rm Let $W, V$ be finite-dimensional real vector spaces of vector fields on a manifold $N$.
We say that they form a {\it quasi-Lie scheme} $S(W,V)$ if the following are satisfied
\begin{enumerate}
\item $W$ is a vector subspace of $V$. \item $W$ is a Lie algebra of vector fields, i.e. $[W,W]\subset W$.
\item $W$ normalises $V$, i.e. $[W,V]\subset V$.
\end{enumerate}
If $V$ is a Lie algebra of vector fields $V$, we call the quasi-Lie scheme $S(V,V)$ simply a {\it Lie scheme}
$S(V)$.}
\end{definition}

\begin{remark}{\rm There is the largest Lie subalgebra we can use as $W$ -- the normalizer of $V$ in $V$. Sometimes, however, it is useful to consider smaller Lie subalgebras $W$.}
\end{remark}
We say that a time-dependent vector field $X$ is in a quasi-Lie scheme $S(W,V)$, and write $X\in S(W,V)$, if
$X$ belongs to $V$ on its domain, i.e. $X_t\in V_{|N^X_t}$. Note that, by definition, the set of all
time-dependent vector fields belonging to $S(W,V)$ depends only on $V$ and the choice of $W$ is irrelevant.

Now, given a quasi-Lie scheme $S(W,V)$ which we will call sometimes simply a scheme, we may consider the group
$\mathcal{G}(W)$ of generalised flows associated with $W$.
\begin{definition} {\rm We call the {\it group of the scheme} $S(W,V)$ the group $\mathcal{G}(W)$
of generalised flows corresponding to the time-dependent vector fields with values in $W$}.
\end{definition}
\begin{proposition}\label{Main} {\bf (Main property of a scheme)}
{\rm Given a scheme $S(W,V)$, a time-dependent vector field $X\in S(W,V)$, and a generalised flow $g\in
\mathcal{G}(W)$, we get that $g_\di X\in S(W,V)$}.
\end{proposition}
The proof is obvious and follows from the fact that if $Y$ is a generalised flow with values in $W$ and $X$
takes values in $V$, then, according to the formula (\ref{e7}), $g^Y_\di X$ takes values in $V$ as well, as
$[W,V]\subset V$ and $V$ is finite-dimensional.

From the last definition we can state the definition of {\rm quasi-Lie system} with respect to a scheme.
\begin{definition} {\rm Given a quasi-Lie scheme $S(W,V)$ and a time-dependent vector field $X\in S(W,V)$,
we say that $X$ is a {\it quasi-Lie system with respect to $S(W,V)$} if there exists a generalised flow $g\in
\mathcal{G}(W)$ and a Lie algebra of vector fields $V_0\subset V$ such that
$$
g_\di X\in S(V_0).
$$}
\end{definition}
We emphasise that if $X$ is a quasi-Lie system with respect to the scheme $S(W,V)$, it automatically admits a
time-dependent superposition rule in the form given by (\ref{e8}). In the next Section, we apply our theory
and illustrate these concepts with examples.

\section{Applications of quasi-Lie schemes}

The above mentioned properties of quasi-Lie schemes and quasi-Lie systems can be used  to investigate some
previously studied systems of differential equations \cite{PK06}-\cite{CRSS04} systematically from this new
perspective.

In this Section, we apply quasi-Lie schemes to study dissipative Milne--Pinney equations, nonlinear oscillators and
Emden differential equations. More precisely, we first apply our theory to dissipative Milne--Pinney equations.
These systems cannot be treated with the theory of Lie systems directly but one can use a quasi-Lie scheme to
transform them into Milne--Pinney equations. These latter equations have been proved to be SODE Lie systems recently
\cite{CRL07a, CLR08c} and this fact enables us to get a time-dependent superposition rules for
dissipative Milne--Pinney equations by means of the superposition rule found for non-dissipative ones.

Next, we analyse nonlinear oscillators. Perelomov studied some of these systems in order to relate them to
other important systems \cite{Pe78}. The cases investigated by Perelomov were selected and obtained by means
of {\it ad hoc} methods. Here we approach some instances treated in \cite{Pe78} in order to explain
Perelomov's work from the point of view of the theory of quasi-Lie schemes. As a result, we show how our
theory provides time-dependent constants of the motion and clarify some points about this work.

We also analyse Emden equations. In this case, we use a quasi-Lie scheme to obtain constants of the motion
for Emden equations whose time-dependent coefficients satisfy certain conditions. Notwithstanding, this is
just a small instance of what can be made by means of our scheme.

Finally, we analyse certain kind of dissipative Mathews--Lakshmanan oscillators \cite{ML74, LR03,CRSS04}. Some kinds of these nonlinear oscillators have been recently investigated from the point of view of Classical and Quantum Mechanics \cite{CRSS04, CL09, CLL09}. Here we just perform a simple application of the theory of quasi--Lie schemes to relate different types of dissipative Mathews-Lakschmanan oscillators to the usual ones.

Let us provide a quasi-Lie scheme to deal with some of the systems investigated in following subsections. Recall that we need to find vector spaces
$W$ and $V$ of vector fields satisfying the three conditions stated in definition \ref{QLscheme}. Consider the vector space $V$
spanned by the linear combinations of the vector fields
\begin{equation}\label{BasisV}
X_1=x\pd{}{v},\quad X_2=x^n\pd{}{v},\quad X_3=v\pd{}{x},\quad X_4=v\pd{}{v},\quad X_5=x\pd{}{x}\,
\end{equation}
on ${\rm T}\mathbb{R}$ and take the vector subspace $W\subset V$ generated by
$$
Y_1=X_4=v\pd{}{v},\quad Y_2=X_1=x\pd{}{v},\quad Y_3=X_5=x\pd{}{x}.
$$
Therefore, $W$ is a solvable Lie algebra of vector fields,
$$[Y_1,Y_2]=-Y_2\,,\quad [Y_1,Y_3]=0\,,\quad [Y_2,Y_3]=-Y_2\,,
$$
and taking into account that
\begin{equation*}
\begin{array}{lll}
\left[Y_1,X_2\right]=-X_2,& \left[Y_1,X_3\right]=X_3, & \left[Y_2,X_2\right]=0,
\\ \left[Y_2,X_3\right]=X_5-X_4, &\left[Y_3,X_2\right]=nX_2,&\left[Y_3,X_3\right]=-X_3,
\end{array}
\end{equation*}
we see that $V$ is invariant under the action of $W$, i.e. $[W,V]\subset V$. In this
way we get the quasi-Lie scheme $S(W,V)$. We stress that the vector space $V$ is not a Lie algebra because the commutator $\left[X_2,X_3\right]$ is
not in $V$. Moreover, there is no Lie algebra of vector fields $V'\supseteq V$ and thus $V$ cannot be related to a Lie scheme.
\medskip

The key tool provided by the scheme $S(W,V)$ is the infinite-dimensional group
$\mathcal{G}(W)$ of generalised flows for the time-dependent vector fields with values in $W$,
i.e. $\alpha_1(t)Y_1+\alpha_2(t)Y_2+\alpha_3(t)Y_3$. The integration of such $t$-dependent vector fields lead to the description of the time-dependent changes of variables associated with $\mathcal{G}(W)$, i.e.

{\footnotesize
\begin{equation*}
\mathcal{G}(W)=\left\{g(\alpha(t),\beta(t),\gamma(t))=\left\{
\begin{aligned}
x&=\gamma(t)x'\\
v&=\alpha(t)v'+\beta(t)x'\,
\end{aligned}\right.\bigg|\alpha(t),\gamma(t)>0,\alpha(0)=\gamma(0)=1,\beta(0)=0\right\}.
\end{equation*}}

\subsection{Dissipative Milne--Pinney equations}
\qquad In this Section, we study the so-called dissipative Milne--Pinney equations. We show that the first-order
ordinary differential equations associated with these second-order ones in the usual way, i.e. by considering
velocities as new variables, are not Lie systems. However, the theory of quasi-Lie schemes can be used to deal
with such first-order systems. Here we provide a scheme which enables us to transform a certain kind of
dissipative Milne--Pinney equations, considered as first-order systems, into some first-order Milne--Pinney equations already
studied by means of the theory of Lie systems \cite{CRL07a}. As a result we get a time-dependent superposition
rule for some of these dissipative Milne--Pinney equations.

Let us state the problem under study. Consider the family of dissipative Milne--Pinney equations of the form
\begin{equation}\label{eq1}
\ddot x=a(t)\dot x+b(t) x+c(t)\frac{1}{x^3}\,.
\end{equation}
We are mainly interested in the case $c(t)\not =0$, so we assume  that $c(t)$ has a constant sign for the
 set of values of $t$ we are considering.

Usually, we associate with such a second-order  differential equation a  system of first-order differential
equations by introducing a new variable  $v$ and relating the differential equation (\ref{eq1}) to the system
of first-order differential equations
\begin{equation}\label{eq2}\left\{
\begin{array}{rcl}
\dot x&=&v,\\
\dot v&=&a(t)v+b(t) x+c(t)\dfrac{1}{x^3}.\\
\end{array}\right.
\end{equation}

In order to verify that the quasi-Lie scheme $S(W,V)$, for the case $n=-3$, can be used to handle the latter system, that is, $X\in S(W,V)$, we have to ensure that the time-dependent
vector field
$$
X_t=v\pd{}{x}+\left(a(t)v+b(t)x+\frac{c(t)}{x^3}\right)\pd{}{v}\,,
$$
whose integral curves are solutions for the system (\ref{eq2}), is such  that $X_t\in V$ for every $t$ in an
open interval in $\mathbb{R}$. In this way, in view of (\ref{BasisV}), we observe that
\begin{equation*}
X_t=a(t)X_4+b(t)X_1+c(t)X_2+X_3,
\end{equation*}
and thus $X\in S(W,V)$. Moreover $V''=\langle X_1,\ldots, X_4\rangle$ is not a Lie algebra of vector fields because  $[X_3,X_2]\notin V''$. Also, there is no finite-dimensional real Lie algebra
$V'$ containing $V''$. Thus, system (\ref{eq2}) is not a Lie system but we can use the quasi-Lie scheme
$S(W,V)$ to investigate it.

Let us consider the infinite-dimensional subgroup of $\mathcal{G}(W)$ given by its time-dependent changes of variables with $\gamma(t)=1$. According to the general theory of quasi-Lie schemes, these time-dependent changes of variables enable
us to transform system (\ref{eq2}) into a new one again describing the integral curves for a time-dependent vector field $X'\in S(W,V)$, that is,
\begin{equation}\label{fineq}
X'_t=a'(t)X_4+b'(t)X_1+c'(t)X_2+d'(t)X_3+e'(t)X_5\,.
\end{equation}
The new coefficients are
\begin{equation*}
\left\{
\begin{aligned}
a'(t)&=a(t)-\beta(t)-\frac{\dot \alpha(t)}{\alpha(t)},\\
b'(t)&=\frac{b(t)}{\alpha(t)}+a(t)\frac{\beta(t)}{\alpha(t)}-\frac{\beta^2(t)}
{\alpha(t)}-\frac{\dot\beta(t)}{\alpha(t)},\\
c'(t)&=\frac{c(t)}{\alpha(t)},\\
d'(t)&=\alpha(t),\\
e'(t)&=\beta(t).\\
\end{aligned}\right.
\end{equation*}
The integral curves for the time-dependent vector field (\ref{fineq}) are solutions of the system
\begin{equation}\label{quasiErmsys}\left\{
\begin{array}{rcl}
\dfrac{dx'}{dt}&=&\beta(t)x'+\alpha(t)v',\cr
\dfrac{dv'}{dt}&=&\left(\dfrac{b(t)}{\alpha(t)}+a(t)\dfrac{\beta(t)}{\alpha(t)}-\dfrac{\beta^2(t)}
{\alpha(t)}-\dfrac{\dot\beta(t)}{\alpha(t)}\right)
x'+\\&+&\left(a(t)-\beta(t)-\dfrac{\dot\alpha(t)}{\alpha(t)}\right)v'+\dfrac{c(t)}{\alpha(t)}\dfrac{1}{x'^3}.
\end{array}\right.
\end{equation}
As it was said in Section \ref{Theory}, we use schemes to transform the corresponding systems of first-order
differential equations into Lie ones. So, in this case, we must find a Lie algebra of vector fields
$V_0\subset V$ and a generalised flow $g\in\mathcal{G}(W)$ such that $g_\di X\in S(V_0)$. This leads to a
system of ordinary differential equations for the functions $\alpha(t)$, $\beta(t)$ and some integrability
conditions about the initial functions $a(t),b(t)$ and $c(t)$ for such a time-dependent change of variables to
exist.

In order to find a proper Lie algebra of vector fields $V_0\subset V$, note that Milne--Pinney equations studied in
\cite{CRL07a} are Lie systems in the family of differential equations defined by systems (\ref{eq2}) and
therefore it is natural to look for the conditions needed to  transform a given system (\ref{eq2}),
described by the time-dependent vector field $X_t$, into one of these first-order Milne--Pinney equations of the form
\begin{equation}\label{Ermsys}\left\{
\begin{array}{rcl}
\dot x&=&f(t)v,\\
\dot v&=&-\omega(t) x+f(t)\dfrac{k}{x^3},\,
\end{array}\right.
\end{equation}
 where $k$ is a constant, i.e. a system describing the integral curves for a time-dependent vector field with
 values in the Lie algebra of vector fields \cite{CRL07a}
$$
V_0=\langle X_3+k\,X_2, X_1, \frac{1}{2}(X_5-X_4)\rangle.
$$
As a result, we get
 that $\beta=0$, $\alpha=f$ and, furthermore, the functions  $\alpha$, $a$ and $c$  must satisfy
\begin{equation}\label{alfaeq}
k \alpha^2=c,\qquad \qquad \dot\alpha-a{\alpha}=0,
\end{equation}
so that $c$ must be of constant sign equal to that of $k$. The second condition is a differential equation for
$\alpha$ and the first one determines $c$ in terms of $\alpha$. Therefore, both conditions lead to a relation
between $c$ and $a$ providing the integrability condition
\begin{equation}\label{IntCond}
c(t)=k\,{\rm exp}\left(2A(t)\right), \quad {\rm and}\quad A(t)=\int a(t)dt,
\end{equation}
and showing, in view of (\ref{quasiErmsys}), (\ref{Ermsys}) and (\ref{alfaeq}),  that
$$
\alpha(t)={\rm exp}\left(A(t)\right) \qquad {\rm and}\qquad \omega(t)=-b(t)\exp\left(-A(t)\right),
$$
where we choose the constants of integration in order to get $\alpha(0)=1$ as required.

Summing up the preceding results, under the integrability condition (\ref{IntCond}), the first-order Milne--Pinney
 equation (\ref{eq2}) can be transformed into the system
\begin{equation*}\left\{
\begin{array}{rcl}
\dfrac{dx'}{dt}&=&{\rm exp}\left(A(t)\right)v',\\
&&\\
\dfrac{dv'}{dt}&=&b(t){\rm exp}\left(-A(t)\right) x'+{\rm exp}\left(A(t)\right)\dfrac{k}{x'^3},\,
\end{array}\right.
\end{equation*}
by means of the time-dependent change of variables
\begin{equation*}g\left({\rm exp}\left(A(t)\right),0,1\right)=\left\{
\begin{array}{rcl}
x'&=&x,\\
v'&=&{\rm exp}\left(A(t)\right)v\,
\end{array}\right..
\end{equation*}
\begin{note} The previous change of variables is a particular instance of the so-called
Liouville transformation \cite{Milson}.
\end{note}

Now, the final Milne--Pinney equation can be rewritten by means of the
time reparametrisation
\begin{equation*}
\tau(t)=\int {\rm exp}\left(A(t)\right)dt,
\end{equation*}
as
\begin{equation*}\left\{
\begin{array}{rl}
\dfrac{dx'}{d\tau}&=v',\\
\dfrac{dv'}{d\tau}&={\rm exp}\left(-2A(t)\right)b(t(\tau))x'+\dfrac{k}{x'^3}.
\end{array}\right.
\end{equation*}
These systems were analysed in \cite{CLR07b} and there it was shown through the theory of Lie systems that
they admit the constant of the motion
\begin{equation*}
I=(\bar xv'-\bar vx')^2+k\left(\frac{\bar x}{x'}\right)^2\,,
\end{equation*}
where $(\bar x,\bar v)$ is a  solution of the system
\begin{equation*}\left\{
\begin{aligned}
\frac{d\bar x}{d\tau}&=\bar v,\\
\frac{d\bar v}{d\tau}&={\rm exp}\left(-2A(t)\right)b(t)\, \bar x\,,
\end{aligned}\right.
\end{equation*}
 which can be written  as a second-order differential equation
\begin{equation*}
\frac{d^2\bar x}{d\tau^2}={\rm exp}\left(-2A(t)\right)b(t)\,\bar x\,.
\end{equation*}
If we invert the time reparametrisation,  we obtain the following differential equation
\begin{equation}\label{SO}
\ddot{\bar x}-a(t)\dot{\bar  x}-b(t)\bar x=0,
\end{equation}
which is the linear differential equation associated with the initial Milne--Pinney equation.

As it was shown in \cite{CRL07a}, we can obtain, by means of the theory of Lie systems, the following
superposition rule
$$
x'=\frac {\sqrt 2} {|\bar x_1\bar v_2-\bar v_1\bar
x_2|}\left(I_2\bar{x}_1^2+I_1\bar{x}_2^2\pm\sqrt{4I_1I_2-k(\bar x_1\bar v_2-\bar v_1\bar v_2)^2}\ \bar x_1\bar
x_2\right)^{1/2}\,,
$$
and as the time-dependent transformation performed does not change the variable $x$, we get the time-dependent
superposition rule
$$
x=\frac {\sqrt 2\alpha(t)} {|\bar x_1\dot {\bar x}_2-\dot{ \bar x}_1\bar
x_2|}\left(I_2\bar{x}_1^2+I_1\bar{x}_2^2\pm\sqrt{4I_1I_2-\frac{k}{\alpha^2(t)}(\bar x_1\dot {\bar x}_2-\dot{
\bar x}_1\bar x_2)^2}\ \bar x_1\bar x_2\right)^{1/2}\,,
$$
in terms of a set of solutions of the second-order linear system (\ref{SO}).

Summing up, the application of our scheme to the family of dissipative Milne--Pinney equations
\begin{equation*}
\ddot x=a(t)\,\dot x+b(t)\,x+ {\rm exp}\left(2\int a(t)\,dt\right)\frac{k}{x^3}
\end{equation*}
shows that it admits a time-dependent superposition principle:
$$
x=\frac {\sqrt 2\alpha(t)} {|y_1\dot y_2-y_2\dot
y_1|}\left(I_2y_1^2+I_1y_2^2\pm\sqrt{4I_1I_2-\frac{k}{\alpha^2(t)}(y_1\dot y_2-y_2\dot y_1)^2}\,  y_1
y_2\right)^{1/2}\,,
$$
in terms of two independent solutions $y_1,y_2$ for the differential equation
\begin{equation*}
\ddot{y }-a(t)\,\dot{y}-b(t)\,y=0.
\end{equation*}

So, we have fully detailed a particular application of the theory of quasi-Lie schemes to dissipative Milne--Pinney equations. As a result, we provide a time-dependent superposition rule for a family of such systems.  Another
paper dealing with such an approach to dissipative Milne--Pinney equations and explaining some of their properties can
be found in \cite{CL08p}.

\subsection{Nonlinear oscillators}
\qquad As a second application of our theory, we use quasi-Lie schemes to deal with a certain kind of nonlinear
oscillators. The main objective of this Section is to explain some properties of some time-dependent
nonlinear oscillators studied by Perelomov in \cite{Pe78}. We also furnish with a, as far as we know, new
constant of the motion for these systems.

Consider the subset of the family of nonlinear oscillators investigated in \cite{Pe78}:
\begin{equation*}
\ddot x=b(t)x+c(t)x^n,\qquad n\neq 0,1\,.
\end{equation*}
The cases $n=0,1$, are omitted because they can be handled with the usual theory of Lie systems. Like in the
above Section, we link the above second-order ordinary differential equation to the first-order system
\begin{equation}\label{NLO2}\left\{
\begin{aligned}
\dot x&=v,            \\
\dot v&=b(t)x+c(t)x^n.\cr
\end{aligned}\right.
\end{equation}

Now, we have to go over whether the solutions of system (\ref{NLO2}) are integral curves for a time-dependent
vector field $X\in S(W,V)$. In order to check out this, we realise that the system (\ref{NLO2})
describes the integral curves for the time-dependent vector field
$$
X_t=v\pd{}{x}+(b(t)x+c(t)x^n)\pd{}{v},
$$
which can be written as
\begin{equation}
X_t=b(t)X_1+c(t)X_2+X_3\,.\label{NLOPLS}
\end{equation}

Note also that  $[X_2,X_3]\notin V$  and not only $V''=\langle X_1,X_2,X_3\rangle$ is not a Lie algebra
of vector fields but also there is no finite-dimensional Lie algebra $V'$ including $V''$. Thus, $X$ cannot be
considered as a Lie system and we conclude that the first-order nonlinear oscillator (\ref{NLO2}) describing integral curves of the time-dependent vector field (\ref{NLOPLS}) (which is not a Lie system) can be described by means of the quasi-Lie scheme $S(W,V)$.

Let us restrict ourselves to analyse those time-dependent changes of variables associated with the generalised flows of $\mathcal{G}(W)$ with $\beta(t)=\dot\gamma(t)$ and $\alpha(t)=1/\gamma(t)$ and apply these transformations
to the system (\ref{NLO2}). The main theorem of the theory of quasi-Lie systems tells us that
$$g(\alpha(t),\beta(t),\gamma(t))_\di X\in S(W,V).$$ Indeed, these time-dependent transformations
lead to the systems
\begin{equation}\label{transformed}
\left\{
\begin{aligned}
\frac{dx'}{dt}&=\frac{1}{\gamma^2(t)}v',\\
\frac{dv'}{dt}&=(\gamma^2(t)b(t)-\ddot\gamma(t)\gamma(t))x'+c(t)\gamma^{n+1}(t)x'^n,\,
\end {aligned}
\right.
\end{equation}
which are related to the second-order differential equations
\begin{equation*}
\gamma^2(t) \ddot x' =-2\gamma(t)\dot\gamma(t)\dot
x'+(\gamma^2(t)b(t)-\ddot\gamma(t)\gamma(t))x'+c(t)\gamma^{n+1}(t)x'^n\,.
\end{equation*}
But the theory of quasi-Lie schemes is based on finding a  generalised flow  $g\in\mathcal{G}(W)$ such
that ${g}_\di X$ becomes a Lie system, i.e. there exists a Lie algebra of vector fields $V_0\subset V$ such
that ${g}_\di X\in S(V_0)$. For instance, we can try to transform a particular instance of the systems
(\ref{transformed}) into a first-order differential equation associated with a nonlinear oscillator with a zero
time-dependent angular frequency, for example, into the first-order system
\begin{equation}\label{sys}
\left\{
\begin{aligned}
\frac{dx'}{dt}&=f(t)v',\\
\frac{dv'}{dt}&=f(t)c_0x'^n\,,
\end {aligned}
\right.
\end{equation}
related to the nonlinear oscillator
$$
\frac{d^2x'}{d\tau^2}=c_0x'^n,
$$
with $d\tau/dt=f(t)$.

The conditions ensuring such a transformation are
\begin{equation}\label{condit2}
\gamma(t)b(t)-\ddot\gamma(t)=0\,,\quad c(t)=c_0\gamma^{-(n+3)}(t),
\end{equation}
with $f(t)=\gamma^{-2}_1(t)$, where $\gamma_1$ is a  non-vanishing particular solution for
$\gamma(t)b(t)-\ddot\gamma(t)=0$. We must emphasise that just particular solutions  with $\gamma_1(0)=1$ and $\dot\gamma_1(0)=0$ are related to generalised flows in $\mathcal{G}(W)$. Nevertheless, any other particular solution can be used also to transform a nonlinear oscillator into a Lie system as we stated. The Lie system (\ref{sys}) is the system associated with
the time-dependent vector field
\begin{equation*}
X_t=\frac1{\gamma^{2}_1(t)}\left(v'\pd{}{x'}+c_0x'^n\pd{}{v'}\right).
\end{equation*}

As a consequence of the standard methods developed for the theory of Lie systems \cite{CLR08c}, we  join two
copies of the above system in order to get the first-integrals
\begin{equation*}
I_i=\frac{1}{2}v_i'^2-\frac{c_0}{n+1}x_i'^{n+1}, \qquad i=1,2,
\end{equation*}
and
\begin{multline*}
I_3=\frac{x'_1}{\sqrt{I_1}}{\rm
Hyp}\left(\frac{1}{n+1},\frac{1}{2},1+\frac{1}{n+1},-\frac{c_0x_1'^{n+1}}{I_1(n+1)}\right)\\-
\frac{x'_2}{\sqrt{I_2}}{\rm
Hyp}\left(\frac{1}{n+1},\frac{1}{2},1+\frac{1}{n+1},-\frac{c_0x_2'^{n+1}}{I_2(n+1)}\right),
\end{multline*}
where ${\rm Hyp}(a,b,c,d)$ denotes the corresponding hypergeometric functions. In terms of the
initial variables these first-integrals for ${g}_\di X$ read

\begin{equation}\label{integral1}
\begin{aligned}
I_i&=\frac{1}{2}(\gamma_1(t)\dot x_i-\dot\gamma_1(t)x_i)^2-\frac{c_0}{\gamma_1^{n+1}(t)(n+1)}x_i^{n+1},\qquad i=1,2,
\end{aligned}
\end{equation}
and
\begin{multline}\label{integral2} I_3=\frac{1}{\gamma_1(t)}\left(\frac{x_1}{\sqrt{I_1}}{\rm
Hyp}\left(\frac{1}{n+1},\frac{1}{2},1+\frac{1}{n+1},-\frac{c_0x_1^{n+1}}{\gamma_1^{n+1}(t)I_1(n+1)}\right)\right.\\
\left. -\frac{x_2}{\sqrt{I_2}}{\rm
Hyp}\left(\frac{1}{n+1},\frac{1}{2},1+\frac{1}{n+1},-\frac{c_0x_2^{n+1}}{\gamma_1^{n+1}(t)I_2(n+1)}\right)\right).
\end{multline}

As a particular application of conditions (\ref{condit2}), we can consider the following example of \cite{Pe78}, where
the time-dependent Hamiltonian
\begin{equation*}
H(t)=\frac{1}{2}p^2+ \frac{\omega^2(t)}{2}x^2+c^2\gamma_1^{-(s+2)}(t)x^s\,,
\end{equation*}
with $\gamma_1$ being such that $\ddot\gamma_1(t)+\omega^2(t)\gamma_1(t)=0$, is studied. The Hamilton
equations for the latter Hamiltonian are
\begin{equation}\label{IQLS}\left\{
\begin{aligned}
\dot x&=p,\\
\dot p&=-sc^2\gamma_1^{-(s+2)}(t)x^{s-1}-\omega^2(t)x,
\end{aligned}\right.
\end{equation}
being associated with the second-order differential equation for the variable $x$ given by
\begin{equation}\label{NLOS}
\ddot x=-sc^2\gamma_1^{-(s+2)}(t)x^{s-1}-\omega^2(t)x.
\end{equation}

The latter differential equations are  particular cases of our Emden equations with
\begin{equation}\label{OurCase}
b(t)=-\omega^2(t)\,,\qquad c(t)=-sc^2\gamma_1^{-(s+2)}(t)\,,\quad n=s-1.
\end{equation}
Notice that here the variable $p$ plays the role of $v$ in our theoretical development. It can be easily verified that
 these coefficients satisfy the conditions (\ref{condit2}).
% \begin{enumerate}
%  \item By assumption, $\omega^2(t)\gamma_1(t)+\ddot\gamma_1(t)=0$. As $\omega^2(t)=-b(t)$, the latter means that
%  $\gamma_1(t)b(t)-\ddot\gamma_1(t)=0$.
% \item If we fix $c_0=-sc^2$, in view of conditions (\ref{OurCase}), we obtain $c(t)=c_0\gamma_1^{-(n+3)}(t)$.
% \end{enumerate}
Therefore, we get that the time-dependent  frequency nonlinear oscillator (\ref{NLOS}) can be transformed into
a new one with zero frequency, i.e.
$$
\frac{d^2x'}{d\tau^2}=-sc^2x'^{s-1},
$$
with
$$
\tau=\int \frac{dt}{\gamma^2_1(t)},
$$
reproducing the result given by Perelomov \cite{Pe78}. The choice of the time-dependent frequencies is such that it
is possible to transform the initial time-dependent nonlinear oscillator into the  final autonomous nonlinear
oscillator. Then, we recover here such frequencies as a result of an integrability condition. Moreover, in
view of the expressions  (\ref{integral1}), (\ref{integral2}) and (\ref{OurCase}), we get a, as far as we know,
new time-dependent constants of the motion for these nonlinear oscillators.

\subsection{The Emden equation}
\qquad In this Section we apply quasi-Lie schemes to Emden equations. These equations appear broadly in the
literature and have many applications, indeed, the review by Wong in 1977 \cite {Wong76a} contains more than one hundred cites. Here
they are analysed  in order to recover, under some integrability conditions, a set of time-dependent constants
of the motion.

The Emden equations we investigate are
\begin{equation}\label{Emden}
\ddot x=a(t)\dot x+b(t)x^n,\quad n\ne 1\,.
\end{equation}
The case $n=1$ is removed because it can be treated directly by means of the theory of Lie systems.

Emden equations are associated with the system of first-order differential equations
\begin{equation}\label{Emdsys}\left\{
\begin{aligned}
\dot x&=v,\\
\dot v&=a(t)v+b(t)x^n.
\end{aligned}\right.
\end{equation}

As in the preceding examples, let us verify that the scheme $S(W,V)$ can be used to handle this system. The system (\ref{Emdsys}) describes the integral curves for the time-dependent vector field given by
$$
X_t=v\pd{}{x}+(a(t)v+b(t)x^n)\pd{}{v},
$$
which, in terms of the basis (\ref{BasisV}) for $V$, reads
$$X_t=a(t)X_4+X_3+b(t)X_2\,,
$$
so that $X\in S(W,V)$. We must remark that, as $[X_3,X_2]\notin V$, there is no Lie algebra of vector fields
containing the vector space  $V''$ spanned by $X_2,X_3,X_4$ and $X_t$ cannot considered as a Lie system.

The time-dependent change of variables induced by a control $g\in\mathcal{G}(W)$ transforms the system
(\ref{Emdsys}) into {\small
\begin{equation}\label{transformed2}\left\{
\begin{aligned}\frac{dx'}{dt}&=\left(\frac{\beta(t)}{\gamma(t)}-\frac{\dot \gamma(t)}{\gamma(t)}\right)x'+
\frac{\alpha(t)}{\gamma(t)}v',\\
\frac{dv'}{dt}&=\left(a(t)-\frac{\beta(t)}{\gamma(t)}-\frac{\dot\alpha(t)}{\alpha(t)}\right)v'+
\left(a(t)\frac{\beta(t)}{\alpha(t)}-\frac{\beta^2(t)}{\alpha(t)\gamma(t)}-\frac{\dot\beta(t)}
{\alpha(t)}+\frac{\beta(t)\dot\gamma(t)}{\alpha(t)\gamma(t)}\right)x'\\&+\frac{b(t)\gamma^n(t)}{\alpha(t)}x'^n.
\end{aligned}\right.
\end{equation}}According to Theorem \ref{Main}, the latter systems describe integral curves of the time-dependent vector
field ${g}_\di X\in S(W,V)$. Now, we must look for a Lie algebra of vector fields
$V_0\subset V$ and a control $g\in\mathcal{G}(W)$ such that ${g}_\di X\in S(V_0)$.

For the sake of simplicity, let us suppose that $\beta(t)=0$. Thus, the system (\ref{transformed2}) leads to

\begin{equation*}\left\{
\begin{aligned}
\frac{dx'}{dt}&=-\frac{\dot \gamma(t)}{\gamma(t)}x'+\frac{\alpha(t)}{\gamma(t)}v',\\
\frac{dv'}{dt}&=\left(a(t)-\frac{\dot\alpha(t)}{\alpha(t)}\right)v'+b(t)\frac{\gamma^n(t)}{\alpha(t)}x'^n.\end{aligned}\right.
\end{equation*}

We impose some conditions ensuring that this differential equation is a solvable Lie system.
 For instance, we want it to be of the form
\begin{equation}\label{finalSys}
\left\{
\begin{aligned}
\frac{dx'}{dt}&=f(t)\left(c_{11}x'+c_{12}v'\right),\\
\frac{dv'}{dt}&=f(t)\left(c_{21}x'^n+c_{22}v'\right),
\end{aligned}\right.
\end{equation}
where the coefficients $c_{ij}$ are constant. Therefore, we get
\begin{equation*}
\begin{array}{rl}
f(t)c_{11}&=-\dfrac{\dot\gamma(t)}{\gamma(t)},\qquad\ \, f(t)c_{12}=\dfrac{\alpha(t)}{\gamma(t)},\\
f(t)c_{21}&=\dfrac{b(t)\gamma^n(t)}{\alpha(t)}, \quad f(t)c_{22}=a(t)-\dfrac{\dot\alpha(t)}{\alpha(t)},
\end{array}
\end{equation*}
that implies
\begin{equation*}
\alpha(t)=-\frac{c_{12}}{c_{11}}\dot\gamma(t)\Longrightarrow -\frac{\ddot\gamma(t)}{\dot\gamma(t)}+a(t)=
-\frac{c_{22}}{c_{11}}\frac{\dot\gamma(t)}{\gamma(t)}.
\end{equation*}
If we fix $c_{22}=1, c_{11}=-1$, $c_{12}=4$ and $c_{21}=-1$, and we define $A(t)=\int a(t)dt$, the values of $\gamma$ and $\alpha$ are
\begin{equation*}
\gamma(t)=\sqrt{2\int\exp\left(A(t)\right)dt},\qquad \alpha(t)=\frac{4}{\gamma(t)}\exp\left(A(t)\right),
\end{equation*}
with appropriately chosen constants of integration to get $\gamma(0)=\alpha(0)=1$ and $g(\alpha(t),0,\gamma(t))\in \mathcal{G}(W)$. Nevertheless, any other particular solutions with different initial conditions can be used also. Now, since
\begin{equation*}
-\frac{b(t)\gamma^n(t)}{\alpha(t)}=\frac 14\frac{\alpha(t)}{\gamma(t)}\,,
\end{equation*}
we see that
\begin{equation*}
-b(t)\gamma^{n+3}(t)=4\exp\left(2A(t)\right)
\end{equation*}
and
\begin{equation*}
b^{-\frac{2}{n+3}}(t)\exp\left(\frac{4 A(t)}{n+3}\right)-2^{\frac{n-1}{n+1}}\int\exp\left(A(t)\right)dt=0,
\end{equation*}
which is equivalent to the expression found in \cite{Le85} if we do not fix the initial conditions for $\gamma(t)$ and $\alpha(t)$.

Now, let us obtain a constant of the motion for (\ref{finalSys}). As $\alpha(t)/\gamma(t)=4\,f(t)$, we get
\begin{equation*}
f(t)=\frac{\exp\left(A(t)\right)}{2\int\exp\left(A(t)\right)dt}.
\end{equation*}
Hence, the system (\ref{finalSys}) admits a constant of the motion in the form
\begin{equation*}
I=-2\,v'^2-\frac{x'^{n+1}}{n+1}+x'v'.
\end{equation*}
If we invert the initial change of variables,
 we reach the following constant of the motion for our initial differential equation,
\begin{equation*}
I'=\left(v^2-\frac{2b(t)}{n+1}x^{n+1}\right)\exp\left(-2A(t)\right)\int\exp\left(A(t)\right)dt-
xv\exp\left(-A(t)\right),
\end{equation*}
which is equivalent to the one found by Sarlet and Bahar in \cite{SB80}.

\subsection{Dissipative Mathews--Lakshmanan oscillators}
In this Section we provide a simple application of the theory of quasi-Lie schemes to investigate the time-dependent dissipative Mathews-Lakshmanan oscillator
\begin{equation}\label{DML}
(1+\lambda x^2)\ddot x-F(t)(1+\lambda x^2)\dot x-(\lambda x)\dot x^2+\omega(t)x=0,\qquad \lambda>0.
\end{equation}
More specifically, we supply some integrability conditions to relate it to the Mathews--Lakshmanan oscillator \cite{ML74,LR03,CRSS04,CRS04}
\begin{equation}\label{ML}
(1+\lambda x^2)\ddot x-(\lambda x)\dot x^2+k x=0,\qquad \lambda>0,
\end{equation}
and by means of such a relation we get a, as far as we know, new time-dependent constant of the motion.

Consider the system of first-order differential equation related to equation (\ref{DML}) in the usual way, i.e.
\begin{equation}\label{FirstML}
\left\{
\begin{aligned}
\dot x&=v,\\
\dot v&=F(t)v+\frac{\lambda x v^2}{1+\lambda x^2}-\omega(t)\frac{x}{1+\lambda x^2},
\end{aligned}\right.
\end{equation}
and determining the integral curves for the time-dependent vector field
$$
X_t=\left(F(t) v+\frac{\lambda x v^2}{1+\lambda x^2}-\omega(t)\frac{x}{1+\lambda x^2}\right)\frac{\partial}{\partial v}+v\frac{\partial}{\partial x}.
$$
Let us provide a scheme to handle the system (\ref{FirstML}). Consider the vector space $V$ spanned by the vector fields
\begin{equation}\label{MLbasis}
X_1=v\frac{\partial}{\partial x}+\frac{\lambda x v^2}{1+\lambda x^2}\frac{\partial}{\partial v},
\quad X_2=\frac{x}{1+\lambda x^2}\frac{\partial}{\partial v},\quad X_3=v\frac{\partial}{\partial v},
\end{equation}
and the linear space $W=\langle X_3\rangle$. The commutator relations
$$
[X_3,X_1]=X_1,\qquad [X_3,X_2]=-X_2,
$$
imply that the linear spaces $W,V$ made up a quasi-Lie scheme $S(W,V)$.  As the time-dependent vector field $X_t$ reads in terms of the basis (\ref{MLbasis})
$$
X_t=F(t)X_3-\omega(t)X_2+X_1,
$$
we get that $X_t\in S(W,V)$.

The integration of $X_3$ shows that
{\footnotesize
\begin{equation*}
\mathcal{G}(W)=\left\{g(\alpha(t))=\left\{
\begin{aligned}
x&=x',\\
v&=\alpha(t)v'.\,
\end{aligned}\right.\bigg|\,\alpha(t)>0,\,\alpha(0)=1\right\},
\end{equation*}}
and the time-dependent changes of variables related to the controls of $\mathcal{G}(W)$ transform the system (\ref{FirstML}) into
\begin{equation*}
\left\{
\begin{aligned}
\dot x'&=\alpha(t)v',\\
\dot v'&=\left(F(t)-\frac{\dot \alpha(t)}{\alpha(t)}\right)v'-\frac{\omega(t)}{\alpha(t)}\frac{x'}{1+\lambda x'^2}+\alpha(t)\frac{\lambda x'v'^2}{1+\lambda x'^2}.
\end{aligned}\right.
\end{equation*}
Suppose that we fix $\dot \alpha-F(t)\alpha=0$. Hence, the latter becomes
\begin{equation*}
\left\{
\begin{aligned}
\dot x'&=\alpha(t)v',\\
\dot v'&=-\frac{\omega(t)}{\alpha(t)}\frac{x'}{1+\lambda x'^2}+\alpha(t)\frac{\lambda x'v'^2}{1+\lambda x'^2}.
\end{aligned}\right.
\end{equation*}
Let us try to search conditions for ensuring the above system to determine the integral curves for a time-dependent vector field of the form $X(t,x)=f(t)X(x)$ with $X\in V$, e.g.
\begin{equation*}
\left\{
\begin{aligned}
\dot x'&=f(t)v',\\
\dot v'&=f(t)\left(\frac{x'}{1+\lambda x'^2}+\frac{\lambda x'v'^2}{1+\lambda x'^2}\right).
\end{aligned}\right.
\end{equation*}
In such a case, $\alpha(t)=f(t)$, $\omega(t)=-\alpha^2(t)$ and therefore $\omega(t)=-\exp\left(2\int F(t)dt\right)$. The time-reparametrization $d\tau=f(t)dt$ transforms the previous system into the autonomous one
\begin{equation*}
\left\{
\begin{aligned}
\frac{dx'}{d\tau}&=v',\\
\frac{dv'}{d\tau}&=\frac{x'}{1+\lambda x'^2}+\frac{\lambda x'v'^2}{1+\lambda x'^2}.
\end{aligned}\right.
\end{equation*}
determining the integral curves for the vector field $X=X_1+X_2$ and related to a Mathews--Lakshmanan oscillator (\ref{ML}) with $k=1$.
The method of characteristics shows after a brief calculation that this system has a first-integral
$$
I(x',v')=\frac{1+\lambda x'^2}{1+\lambda v'^2},
$$
that reads in terms of the initial variables and the time as a time-dependent constant of the motion
$$
I(t,x,v)=\frac{\alpha^2(t)+\lambda \alpha^2(t)x^2}{\alpha^2(t)+\lambda v^2},
$$
for the time-dependent dissipative Mathews--Lakshmanan oscillator (\ref{DML}) getting a, as far as we know, new $t$-dependent constant of the motion.
\section{Conclusions and Outlook}
\qquad We develop the theory of quasi-Lie schemes as a generalization of the theory of Lie systems and we
prove some of their fundamental properties and find applications. In particular, we recover a time-dependent
superposition rule for a family of dissipative Milne--Pinney equations. This result, which can be found in
\cite{Re99}, is seen here from a new perspective as a result of systematic treatment of the family of
dissipative Milne--Pinney equations admitting  such a time-dependent superposition rule.

Additionally, we explain from a geometric point of view some transformation properties of time-dependent
nonlinear oscillators. More precisely, we provide a geometrical understanding for some results of the
Perelomov's paper \cite{Pe78}. Moreover, quasi-Lie approach allows us to investigate quantum nonlinear
Hamiltonians and supply an explanation of the transformation properties for the quantum analogue of this
physical model \cite{CL09}. Finally, we also recover time-dependent constants of the motion for Emden equations and certain new dissipative time-dependent Mathews--Lakshmanan oscillator.

We hope that this shows that the theory of quasi-Lie schemes can be viewed as a good approach to study many
interesting systems of differential equations and quantum Hamiltonians from the same geometric viewpoint. We
follow this idea in some works under development \cite{CLL09,CL08p}.

Finally, the here developed examples prove that the set of time-dependent changes of variables allowing us to transform a differential equation
in a scheme into a Lie system can be broader than that of those detailed for the group of such a scheme. We have already found an explanation for this fact which will be included in next works within the theory of quasi-Lie schemes \cite{CLL09}.

\section*{Acknowledgements}
Partial financial support by research projects MTM2006-10531 and E24/1 (DGA) and by the Polish Ministry of
Science and Higher Education under the grant No. N201 005 31/0115 are acknowledged. JdL also acknowledge
 a F.P.U. grant from  Ministerio de Educaci\'on y Ciencia.

\end{document}